\documentstyle[fleqn,12pt,english]{article}
\def\beq{\begin{equation}}
\def\eeq{\end{equation}}
\def\0{\otimes}
\def\6{\langle}
\def\9{\rangle}
\def\half{\mbox{$1\over2$}}
\def\sfA{{\sf A}\ }
\def\sfB{{\sf B}\ }

\begin{document}
\begin{center}
{\large{\bf Lorentz transformations of open systems}}\\[10mm]
ASHER PERES and DANIEL R. TERNO\\[8mm]
Department of Physics, Technion---Israel Institute of
Technology, 32000 Haifa, Israel\end{center}\vskip15mm

\noindent{\bf Abstract. }\ We consider open dynamical systems, subject
to external interventions by agents that are not completely described
by the theory (classical or quantal). These interventions are localized
in regions that are relatively spacelike. Under these circumstances,
no relativistic transformation law exists that relates the descriptions
of the physical system by observers in relative motion. Still, physical
laws are the same in all Lorentz frames.
\vskip15mm

\noindent{\bf1. \ A tale of two magicians}

Many years ago, some time in the twentieth century, there were two
itinerant magicians (probably Gypsies) who went from village to village
and entertained people with sleight of hand tricks. One of them called
himself a quantum magician. He claimed to have super\-natural power,
to use super\-symmetric particles and to send quantum information with
super\-luminal velocity. The crowds were amazed by his super\-natural
legerdemain.

A few days later, came the second magician. That one wanted to be
an educator. He showed how to mimic quantum uncertainties with
a mundane deck of cards~\cite{kirk}. He indeed repeated every trick
of the quantum magician, and he then explained how to perform it,
using only the ordinary laws of classical mechanics. People were
dismayed. They realized that they had been fooled: in the preceding
shows they had not seen anything supernatural. No one liked the second
magician, and no one thanked him for his lessons.

For instance, the quantum magician produced two spin~\half\ particles in
a singlet state, far away from each other. His assistant Alice measured
one of the spin components, and {\it instantaneously\/} the wave
function collapsed all over spacetime. All over spacetime? This makes no
sense. A wave function does not live in spacetime, but in a {\it Hilbert
space\/}; and in a Hilbert space, the notion of velocity does not exist.

The second magician just took a piece of ordinary matter at rest. He
made it explode into two fragments which carried opposite angular
momenta, whose directions were unpredictable and had an isotropic
Liouville distribution.  The magician's assistant, called Bob,
measured the angular momentum of one of the fragments and, lo and
behold, the Liouville distribution instantaneously collapsed all over
spacetime. All over spacetime? Of course not. Liouville distributions
do not live in spacetime, but in {\it phase space\/}. Phase space has a
symplectic structure, and does not admit the notion of velocity.\bigskip

\noindent{\bf2. \ Two quantum particles}

The twentieth century witnessed two revolutions in our conception of
nature. The first one was relativity theory: Einstein found that
simultaneity had no absolute meaning and that distant events might have
different time orderings when referred to observers in relative motion.
Einstein's theory elicited strong opposition when it was proposed, but
is generally accepted by now. On the other hand, the revolution caused
by quantum theory still produces uneasy feelings among some physicists.
Einstein himself was puzzled by what seemed to be instantaneous
transmission of quantum information.  In his autobiography
\cite{schilpp} he used terms such as ``telepathically'' (p.~85) and
``spook'' (p.~683).

Since then, the ``peaceful coexistence'' \cite{shimony} of special
relativity and quantum measurement theory was the subject of numerous
theoretical articles \cite{WZ}. The first attempts to discuss the
compatibility of both theories were rather naive: the only relativistic
feature that was used was the existence of an upper bound, the velocity
of light, on the speed of propagation of physical effects. This mere
limitation does not do justice to the fundamentally new concepts
introduced by Einstein's relativity (one could as well imagine
information theory limited by the speed of sound, or that of the postal
service).

A more subtle issue, that came under scrutiny in later years, is
that a physical situation involving several observers in relative
motion cannot be described by wave functions with a relativistic
transformation law \cite{qt,NYAS}. This is true even if we allow that
law to be nonlocal. (Apparatuses with parts in relative motion have been
the subject of recent experiments \cite{geneva1,geneva2}). The root of
the difficulty we have to transform quantum expressions from
one Lorentz frame to the other is that the process called ``quantum
measurement'' is an intervention in the quantum dynamics by an
``exosystem'' \cite{finkel}, namely by an apparatus which is not
completely described by the quantum formalism \cite{interv}.

Consider the following simple example, illustrated by Fig.~1 which
shows the world lines of two observers (``Alice'' and ``Bob'')
who are receding from each other. Two spin~$1\over2$ particles are
initially prepared in identical states, $|\sigma_x\9=1$, far away
from each other, in the frame of reference of the quantum magician,
where they are at rest. Note that these particles are {\it not\/}
entangled: the problem we are discussing has no relation whatsoever
to quantum entanglement. Alice and Bob, as seen in that frame,
move in opposite $\pm x$ directions. They simultaneously measure
the values of $\sigma_y$ of their particles, in regions \sfA and
\sfB respectively. Actually these are not quite the same variables
$\sigma_y$, because Alice and Bob are in relative motion in the $x$
direction and there is a transformation law between their operators
$\sigma_y$~\cite{thomas}. However, this detail is irrelevant in the
present discussion. The main point is that their results, $\pm1$,
are equiprobable and unpredictable, except statistically.

For example, suppose that Alice finds $\sigma_{yA}=1$ and Bob finds
$\sigma_{yB}=-1$. Then at time $t_A=0$ (in Alice's frame), namely
after Alice performed her measurement but before Bob performed his,
the state of the physical system is described by her as

\beq |\psi(t_A=0)\9=|(\sigma_{yA}=1)\9\otimes|(\sigma_{xB}=1)\9.\eeq
(Alice knows that $\sigma_{xB}=1$ because she is cognizant of the
preparation procedure of the two particles.) Likewise, the state at
time $t_B=0$ is described by Bob as

\beq |\psi(t_B=0)\9=|(\sigma_{xA}=1)\9\otimes|(\sigma_{yB}=-1)\9.\eeq
No relativistic linear transformation can convert these expressions
into one another, in a way that would be valid for {\it all\/} possible
results $\sigma_y=\pm1$.\bigskip

\noindent{\bf3. \ Two classical particles}

We shall now show that a similar situation arises for a classical
system whose state is given in any Lorentz frame by a Liouville
function~\cite{balescu}. Recall that a Liouville function expresses our
probabilistic description of a physical system---what we can predict
before we perform an actual observation---just as the quantum wave
function is a mathematical expression used for computing probabilities
of events~\cite{realism}.

To avoid any misunderstanding, we emphasize that there is no consistent
relativistic statistical mechanics for $N$ interacting particles, with a
$6N$-dimensional phase space defined by the canonical coordinates ${\bf
p}_n$ and ${\bf q}_n$ ($n=1,\ldots,N$). Any relativistic interaction
must be mediated by {\it fields\/}, having an infinity of degrees of
freedom. (The same is true in quantum mechanics: we need quantum field
theory to have interactions compatible with special relativity.) A
complete Liouville function, or rather Liouville functional, must
therefore contain not only all the canonical variables ${\bf p}_n$
and ${\bf q}_n$, but also all the fields. However, once this Liouville
functional is known (in principle), we can define from it a reduced
Liouville function, by integrating the functional over all the degrees
of freedom of the fields. The result is a function of ${\bf p}_n$ and
${\bf q}_n$ only (just as we have reduced density matrices in quantum
theory). The time evolution of such reduced Liouville functions
cannot be obtained directly from canonical Hamiltonian dynamics,
without explicitly mentioning the fields, yet these functions are well
defined in any Lorentz frame, and their relativistic transformation
is unambiguous~\cite{balescu}.

As a simple example, consider two equal point masses $m$, identically
prepared by the second magician, far away from each other. These masses
are restricted to move along straight segments with fixed positions on
the $x$-axis in the magician's frame. In that frame, the masses move
with constant velocity and they bounce elastically when they reach the
extremities of their segments. Their momenta thus are $\pm p$, and
their energy is $E_0=(m^2+p^2)^{1/2}$.  At some arbitrary time $t$ in
the magician's frame, each mass receives a kick $k$, due to an external
agent, so that its momentum becomes $k\pm p$. Its energy becomes

\beq E_\pm=\sqrt{m^2+(k\pm p)^2}. \eeq

As before, Alice and Bob have opposite velocities with respect to
the above inertial frame, and they wish to describe the dynamical
evolution of the physical system in terms of their times, $t_A$ and
$t_B$, respectively. The problem is that they know only statistically
the sign of $\pm p$ before the kick, and therefore the correct sign
to use in Eq.~(\theequation).

Let us assume that the kicks $k$ occurred in spacetime regions \sfA
and \sfB like those in Fig.~1. Then, according to Alice, the situation
at time $t_A=0$ is the following: the particle that was kicked in
region \sfA has equal probabilities to have energy $E_1=E_\pm$, and
the other particle still has energy $E_0$ with certainty. (These are the
values of the energies in the magician's frame. They may be transformed
to Alice's frame if we wish to do so.) The Liouville function, averaged
over all the variables except the energies of the two particles, is
concentrated in two points as shown in Fig.~2(a). If Alice actually
measures $E_1$, one of the two points disappears. Bob's description,
on the other hand, is given by Fig.~2(b). Likewise, if he measures
$E_2$, one of the two points of Fig.~2(b) disappears. No Lorentz
transformation of the Liouville function~\cite{balescu} can relate
these different descriptions.

Note that for spacelike planes that intersect the past light-cones of
events \sfA and \sfB, the Liouville function is concentrated at a
single point in the $E_1E_2$ plane, namely $E_1=E_2=E_0$. For those
intersecting the future light-cones of both events, the Liouville
function has a support consisting of four points if no measurement is
performed, because both $E_1$ and $E_2$ can have values $E_\pm$. It
is only for spacelike planes that intersect the future light cone of
one of these regions, and the past light cone of the other one, that
we get two incompatible descriptions, if measurements are performed
at times $t_A$ and $t_B$, respectively.\bigskip

\noindent{\bf4. \ Open systems}

The purpose of this article was to show that similar problems (or
``paradoxes'') occur in classical and quantum relativistic dynamics,
if there are incompletely described physical agents for which we have
only probabilistic data. This is a general property of {\it open\/}
physical systems, irrespective of the details of their dynamics.

In real life, there are no closed physical systems. We may, if we wish,
imagine that closed systems exist, but since there is no communication
with them, their properties are irrelevant. Still, when we learn (or
teach) physics, physical laws are usually formulated in a language
appropriate to closed systems. For example, Maxwell's equations are
written in terms of vectors {\bf E} and {\bf B}. The components of these
vectors have no objective physical meaning. Their values depends on the
choice of the coordinate system. On the other hand, it is impossible to
formulate the dynamical laws in terms of the scalar quantities ${\bf
E^2-B^2}$ and ${\bf E\cdot B}$. We need {\bf E} and {\bf B} explicitly.
Therefore, a {\it complete\/} description would have to include not only
the components of the electro\-magnetic field, but also the material
realization of the coordinate system used for defining these vector
components. When we say that Maxwell's equations are rotationally
invariant, this actually means that if the material rods that serve us
as coordinate axes are rigidly rotated, then the components of {\bf E}
and {\bf B} have to be replaced by appropriate linear combinations,
in such a way that the equations have the same appearance, in terms
of the new coordinates and new field components.

Moreover, to give a physical meaning to the symbols {\bf E} and
{\bf B}, we should specify how their numerical values are actually
measured, say by means of electrometers and magnetometers, or other
suitable instruments. Now, normally we don't want to be bothered by
how the field components are actually measured, even less by how the
spacetime coordinates are materialized. These technical details are,
after all, irrelevant to electro\-magnetic theory. It is tacitly assumed
that coordinates are well defined and that precise measuring instruments
exist at every spacetime point, and we can proceed with the calculations
without having to think about them. However, as shown in this paper,
what is acceptable and convenient in a deterministic classical theory
becomes problematic if stochastic features are present in the dynamical
evolution, as it happens in quantum phenomena and also in classical
statistical mechanics.  The ``spooky'' long range quantum correlations
that were mentioned by Einstein~\cite{schilpp} also appear in
classical systems if the latter are incompletely specified and treated
statistically by means of Liouville functions (correlations have a
meaning only in statistical analyses).

The crucial feature common to both models discussed above is the
intervention of ``exosystems''~\cite{finkel} (the measuring apparatus
and the $k$-kicker) that are not described by the dynamical formalism of
the ``endosystem'' under consideration, and induce a stochastic behavior
of the latter. The point is that a physical system is ``open'' when
parts of the universe are excluded from its description. At time
$t_A=0$, one of the exosystems has intervened at \sfA but the other
one, that did not intervene as yet, is irrelevant and need not be
excluded. Conversely, at time $t_B=0$ it is the first exosystem that
is irrelevant and need not be excluded. Therefore at times $t_A=0$ and
$t_B=0$ different parts of the universe are excluded, the two systems
are different, and this explains why no Lorentz transformation exists
that relates them. This argument can also be rephrased in terms of
quantum contextuality~\cite{khr}.

It is noteworthy that another causality problem is related to stochastic
exosystems: causal loops (that is, hypothetical effects of future
events on past ones) are dynamically inconsistent if, and only if,
we allow exophysical agents to influence the evolution of physical
systems~\cite{schulman,gatlin}. For closed systems, fully described
by the theory, causal loops lead to no inconsistency and there is
effectively no difference between past and future.\bigskip

\noindent{\bf5. \ Concluding remarks}

Just as the second magician, we don't expect anyone to thank us for
dispelling some of the ``paradoxes'' of the quantum folklore. It is
unfortunate that these fallacious paradoxes are often invoked as a
source of dissatisfaction with the axioms of quantum mechanics. We have
shown that similar features also appear in classical theory. They are
solely due to the introduction of exosystems, which are necessary for
the interpretation of the theory, but are not described by the latter
\cite{pz}. 

One may be tempted to say that it is stochasticity, not quantum theory,
that requires ``peaceful coexistence''~\cite{shimony} with special
relativity. However, there is more in quantum theory than just the
existence of probabilistic data. A true quantum magician, using Bell's
theorem~\cite{bell}, would be able to display long range correlations
that cannot be reproduced by a classical magician. Bell's theorem
implies severe restrictions on our ability to have together realism
and locality. The essential point we wanted to make in this paper is
that in discussions of these fundamental issues, the fact that physical
systems are open should never be forgotten.

\bigskip

\noindent{\bf Acknowledgments}

We thank Larry Schulman and Abner Shimony for helpful
comments. Work by AP was supported by the Gerard Swope Fund and the
Fund for Encouragement of Research. DRT was supported by a grant from
the Technion Graduate School.

\vfill

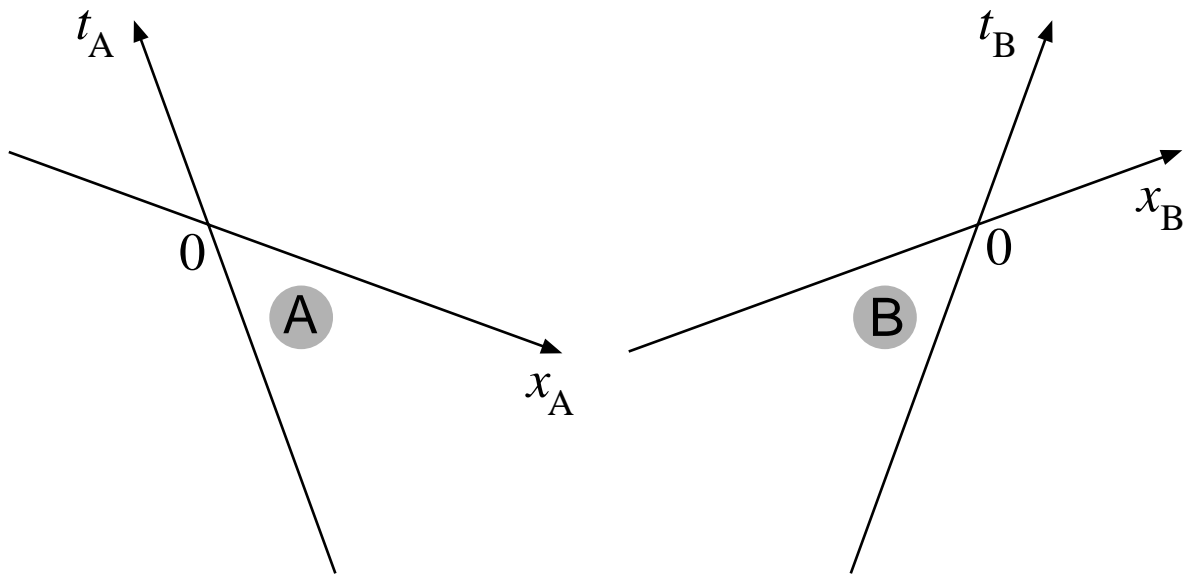
FIG. 1. \ Events that occur simultaneously (in the magicians' frame)
in regions \sfA and \sfB have different time orderings when described
in the Lorentz frames of Alice and Bob.\bigskip

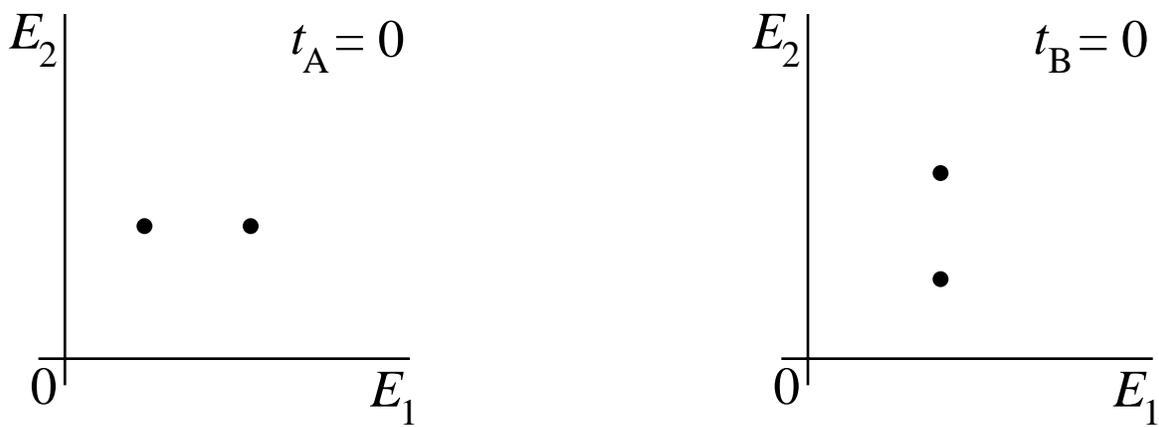
FIG. 2. \ Liouville functions, projected on the energy plane $E_1E_2$,
at times $t_A=0$ and $t_B=0$, respectively, if no measurements are
performed to resolve the indeterminacy. The dots indicate values of
$E_1$ and $E_2$ for which the value of the Liouville function does
not vanish. 

\begin{thebibliography}{99}
\bibitem{realism}Some physicists hold the opinion that wave functions
are physical entities, not only mathematical tools for computing
probabilities.
\end{thebibliography}
\end{document}